\documentstyle[12pt]{article}

\input{tcilatex}
\begin{document}

\title{Mesoscopic circuits with charge discreteness: quantum transmission lines}
\author{J. C. Flores}
\date{Universidad de Tarapac\'a, Departamento de F\'\i sica, Casilla 7-D, Arica,
Chile}
\maketitle

\baselineskip=16pt

We propose a quantum Hamiltonian for a transmission line with charge
discreteness. The periodic line is composed of an inductance and a
capacitance per cell. In every cell the charge operator satisfies a
nonlinear equation of motion because of the discreteness of the charge. In
the basis of one-energy per site, the spectrum can be calculated explicitly.
We consider briefly the incorporation of electrical resistance in the line.

\[
\]

\-PACS Numbers: 73.23.-b ; 73.21.-b ; 73.63.--b ; 07.50.Ek

Will be published in Phys.Rev.B

\newpage

\begin{section}*{I. Introduction: LC quantum circuits with charge discreteness}
\end{section}

Nowadays technological miniaturization of circuits is increasing and their
mesoscopic aspects become more relevant. Recently, a theory for mesoscopic
circuits was proposed by Li and Chen [1] where charge discreteness was
considered explicitly. This is very related to miniaturization since the
number of charges in these systems is expected to be more and more reduced.
In the $LC$ circuit studied in reference [1], the Hamiltonian operator was
given by 
\begin{equation}
\widehat{H}=\widehat{T}+\widehat{V},
\end{equation}
where $V(\widehat{Q})=\widehat{Q}^{2}/2C,$ with $\widehat{Q}$ the charge
operator, is the electrical energy in the capacitance $C$ and the magnetic
energy operator term $\widehat{T}$, related to the inductance $L$, was given
by 
\begin{equation}
\widehat{T}=\frac{2\hbar ^{2}}{Lq_{e}^{2}}\left( sin^{2}\left( \frac{Lq_{e}}{%
2\hbar }k\right) \right) ,
\end{equation}
in $k$-representation (or pseudo-current representation, $0<k<2\pi $). The
above operator has some resemblance with a mechanical kinetic operator in
the limit of small $k$. In (2) the constant $q_{e}$ represents the
elementary charge in the electric system. In this representation, the charge
operator $\widehat{Q}$ is given by 
\begin{equation}
\widehat{Q}=\frac{i\hbar }{L}\frac{\partial }{\partial k},
\end{equation}
and the corresponding eigenfunctions are $e^{inLq_{e}k/\hbar }$ with
discrete eigenvalues $nq_{e}$ where $n$ is an integer. Since the current
operator $\widehat{I}$ is formally obtained from $\widehat{I}=(\frac{1}{%
i\hbar })[\widehat{H},\widehat{Q}]$ then, in the $k$-representation, it
becomes 
\begin{equation}
\widehat{I}=\frac{\hbar }{Lq_{e}}sin(\frac{Lq_{e}}{\hbar }k).
\end{equation}

In the limit $q_{e}\to 0$ all these operators become the usual ones
associated with a quantum $LC$ circuit with continuous charge [3]. Moreover,
the current operator (4) is bounded with extrema values $\pm \frac{\hbar }{%
Lq_{e}}$. After reference [1], the above Hamiltonian describes phenomena
like persistent current, Coulomb blockage and others.

As pointed-out in [2]{}, the above set of operators defines an algebra with
some similitude to this related to space-time discreteness [4,5,6]. In fact,
the charge-current commutator ${[}\widehat{Q}{,}\widehat{I}{]}$, usually
proportional to the identity, becomes modified with a kinetic term which is
zero when $q_{e}=0$. So, space and charge discreteness could be described
with the same mathematical tools.

On the other hand, dissipation is a subject very related to electric
circuits. This phenomenon defines a decoherence-time related to mesoscopic
aspect. Dissipation can be considered in many forms, the usual one is to
connect the system to a bath with many degree of freedom, and with some
assumptions respect to the decaying rate [{7}]. Phenomenological $RLC$
circuits can be also considered using Caldirola-Kanai theory [8,9,10] in a
direct way [{11]. }

In this paper we are interested in a quantum transmission line. Such
transmission lines are usually used in mesoscopic physics. For instance, in
reference [12], a quantum transmission line with continuous charge was
considered and connected to a metal ring. It was quantized and used as
environment to study zero-point fluctuations influence on a metal ring.
Particularly, we shall consider a quantum transmission line with charge
discreteness. For this purpose we shall use the ideas discussed above. In
this way, we consider a periodic transmission line composed of cells. Every
cell has an inductance $L$ and a capacitance $C$.

In section II, we consider briefly the classical transmission line which
will be quantized by standard procedure. In section III, we present the
quantized Hamiltonian which contains explicitly charge discreteness. Also in
that section, we write the motion equation of the charge in the line using
Heinsenberg equation of motion. In section IV, in the one-energy per site
approximation, we find the spectral properties of the systems{\it .} In
section V, we study briefly the incorporation of electrical resistance in
the quantum transmission line with charge discreteness. In the last section,
conclusions are touched.

\[
\]

\begin{section}*{II.  Classical transmission line}
\end{section}

In this section we shall consider a periodic transmission line composed in
every cell of an inductance $L$ and a capacitance $C$. Classically the
evolution equation for the continuous charge $Q_{l}$, at cell $l$ ($l\in Z$)
in the transmission line, is given by the expression 
\begin{equation}
L\frac{d^{2}}{dt^{2}}Q_{l}=\frac{1}{C}\left( Q_{l+1}+Q_{l-1}-2Q_{l}\right) ,
\end{equation}
which can be obtained from the classical Hamiltonian given by 
\begin{equation}
H_{clas}=\sum_{l}\frac{\phi _{l}^{2}}{2L}+\frac{1}{2C}\left(
Q_{l+1}-Q_{l}\right) ^{2},
\end{equation}
where the variable $\phi _{l}$ corresponds to the magnetic flux in the
inductance at position $l$ and it is proportional to the classical current
in the cell. It is explicitly given by

\begin{equation}
\phi _{l}=L\frac{dQ_{l}}{dt}.
\end{equation}

The integer $l$ in (5) represents the index of the cell at position $l$. The
first term depending on the current in (6) is the equivalent to the kinetic
energy of a mechanical system and is related to the stored magnetic energy
in the inductance. In the quadratic potential, the crossed term ($%
Q_{l+1}Q_{l}/C$) represents the interaction term between cells. In fact, for
continuous variables, the above Hamiltonian is equivalent to this one of
mechanical vibrations and could be quantized directly. By canonical
transformation, the classical Hamiltonian (6) could be transformed to normal
modes and its spectral frequencies are well-known. For us, the important
fact is that the crossed term in (6) defines the interaction between two
consecutive sites or cells. This interaction term will be preserved in the
quantization process.

\[
\]

\begin{section}*{III. A Hamiltonian for quantum transmission lines with charge discreteness}
\end{section}

Due to charge discreteness in the quantum case, the structural changes are
only expected to occur in the kinetic part of (6) where the usual quadratic
term is transformed in a trigonometric function (section I). Then, the
quantum transmission line with charge discreteness can be quantized
directly. So, from the classical Hamiltonian (6), and equations (2) and (3),
the quantum Hamiltonian for the transmission line with charge discreteness is

\begin{equation}
\widehat{H}=\sum_{l}\left\{ \frac{2\hbar ^{2}}{Lq_{e}^{2}}\sin ^{2}\left( 
\frac{Lq_{e}}{2\hbar }k_{l}\right) -\frac{\hbar ^{2}}{2L^{2}C}\left( \frac{%
\partial }{\partial k_{l+1}}-\frac{\partial }{\partial k_{l}}\right)
^{2}\right\} ,
\end{equation}
where the quantity $k_{l}$ corresponds to the pseudo-current in the cell at
position $l$ and varies between $0<k_{l}<2\pi $. As pointed before, the
kinetic part is not quadratic in $k_{l}$, which is very related to the
charge discreteness assumptions ($q_{e}\neq 0$). The study of the
Hamiltonian (8) is the purpose of this work. Remark that in the limit $%
q_{e}\rightarrow 0$ the kinetic part becomes proportional to $k_{l}^{2}$
with formal similitude to the case of mechanical vibrations (phonons). This
corresponds to the continuous charge case and the name {\it circuitons} for
these propagating modes is appropriate. Nevertheless, {\it circuitons} are
the limit case with zero charge discreteness $q_{e}\rightarrow 0$. In a
general context, since the above Hamiltonian describes the quantization of
the line and the quantization of the charge we shall call {\it cirquitons}
the normal modes (propagating modes) of the above Hamiltonian. This
appellation seems appropriate because it remembers: (i) the quantization
process and (ii) the discreteness of the charge ($q_{e}$ ). The existence of
these propagating modes is ensured since the Hamiltonian (8) is invariant
under (discrete) spatial traslation.

The motion equation related to the quantum transmission line can be found
with the usual quantum mechanical evolution rules (Heisenberg equations).
The pseudo-current $k_{l}$ has associated the canonical conjugate operator
of the charge, namely, the operator $\widehat{K}_{l}$ with eigenvalues $%
k_{l} $. It satisfies the canonical commutation relation :

\begin{equation}
L[\widehat{Q}_{l},\widehat{K}_{l}]=i\hbar .
\end{equation}

As function of this operator, the Hamiltonian of the transmission line
becomes

\begin{equation}
\widehat{H}=\sum_{l}\left\{ \frac{2\hbar ^{2}}{Lq_{e}^{2}}\sin ^{2}\left( 
\frac{Lq_{e}}{2\hbar }\widehat{K}_{l}\right) +\frac{1}{2C}\left( \widehat{Q}%
_{l+1}-\widehat{Q}_{l}\right) ^{2}\right\} ,
\end{equation}
and the evolution equation for the charge operator in the Heinsenberg
representation $\frac{d}{dt}\widehat{Q}_{l}=\frac{i}{\hbar }[\widehat{H},%
\widehat{Q}_{l}]$ can be computed explicitly:

\begin{equation}
\frac{d}{dt}\widehat{Q}_{l}=\frac{\hbar }{q_{e}L}\sin \left( \frac{Lq_{e}}{%
\hbar }\widehat{K}_{l}\right) .
\end{equation}
Which is similar to the nonlinear expression (4) for every cell. In fact ,
(11) defines the current operator and it is bounded like to the case
mentioned in the section I. The motion equation for the pseudo-current
operator $\frac{d}{dt}\widehat{K}_{l}=\frac{i}{\hbar }[\widehat{H},\widehat{K%
}_{l}]$ is

\begin{equation}
\frac{d}{dt}\widehat{K}_{l}=\frac{1}{CL}\left( \widehat{Q}_{l+1}+\widehat{Q}%
_{l-1}-2\widehat{Q}_{l}\right) .
\end{equation}

In equations (11) and (12), we notice that the formal limit $%
q_{e}\rightarrow 0$ gives the usual linear motion equation for transmission
lines with similitude to the classical one (5). Nevertheless, in the general
case, the evolution equation is nonlinear.

\[
\]

\begin{section}*{IV.  One-energy per site:  excitations on the transmission line}
\end{section}

The study of the spectral properties of the Hamiltonian (8), or (10), is
difficult because of the nonlinear term associated with the magnetic energy.
Without charge discreteness ($q_{e}\rightarrow 0$), this operator is
quadratic and the usual normal modes technique could be used [13]. In this
section we solve the spectral properties on the vector-basis of the system
without interaction. We shall find a particular spectral solution in the
basis of one-energy per site or cell.

The Hamiltonian (10) can be written as

\begin{equation}
\widehat{H}=\sum_{l}\widehat{H}_{l}-\frac{1}{C}\widehat{Q}_{l+1}\widehat{Q}%
_{l},
\end{equation}
where the site-Hamiltonian $\widehat{H}_{l}$ corresponds to this one of a $%
LC $ circuit (1) with capacitance $C/2$. Since the total Hilbert space,
where the Hamiltonian (13) acts, is the direct product of the spaces
associated to every site $l$, then we consider the sub-basis$\left\{ \mid
l\rangle ,l\in Z\right\} $ where every element $\mid l\rangle $ is an
eigenstate of the Hamiltonian $\widehat{H}_{l}$ with energy $V$. Namely,

\begin{equation}
\widehat{H}_{l}\mid l\rangle =V\mid l\rangle .
\end{equation}
Since $\langle n$ $\mid l\rangle =\delta _{l,n}$ and the operator $\widehat{Q%
}_{l}$ acts only in the site $l$, the complete Hamiltonian $\widehat{H}$ is
Hermitian and becomes tri-diagonal in this basis. Explicitly, it is given by
the matrix elements:

\begin{equation}
\langle l\mid \widehat{H}\mid l\rangle =V;\text{ and }\langle l\mid \widehat{%
H}\mid l+1\rangle =-\alpha \frac{q_{e}^{2}}{C},
\end{equation}
where $\alpha $ is a dimensionless constant which we keep as one ($\alpha =1$%
). Formally, this can be carried-out by an adequate normalization. Noticed
that the off-diagonal terms in (15) are related to the interacting term in
(13), namely, $\frac{1}{C}\widehat{Q}_{l+1}\widehat{Q}_{l}$ and then it was
expected to be proportional to $q_{e}^{2}$.

The $LC$ energy stored in a cell is now spread into a band due to charge
interaction with the two neighboring cells. The spectrum of the tri-diagonal
Hamiltonian is well-known and corresponds to the so-called tight-binding
approximations in Solid State Physics. In fact, assuming a general state of
the form $\left| \psi \right\rangle =\sum \psi _{l}\mid l\rangle $, the
Schr\"{o}dinger equation related to the tri-diagonal Hamiltonian (15)
becomes:

\begin{equation}
E\psi _{l}=V\psi _{l}-\frac{q_{e}^{2}}{C}\left( \psi _{l+1}+\psi
_{l-1}\right) ,
\end{equation}
with $E$ the energy. The eigenstates are of the form $\psi _{l}=e^{i\theta
l},$ where the phase $\theta $ is a real number ($0<\theta <2\pi $), and the
spectrum becomes

\begin{equation}
E=V-\frac{2q_{e}^{2}}{C}\cos \theta ,
\end{equation}
where we have considered an infinite transmission line (i.e. $\theta $ is a
continuous variable). The above expression defines the density of states and
the thermodynamics properties of these lines could be calculated.

The spectrum (17) defines an ensemble of quantum excitations on the
transmission line, nevertheless, they are not the more general because they
are only defined in the sub-space spanned by $\left\{ \mid l\rangle ,l\in
Z\right\} $ in the complete Hilbert space. For instance, others kind of
excitations could be found if we consider two energies by site, or different
energies by site.

\[
\]

\begin{section}*{V. Transmission lines with resistance}
\end{section}

The Hamiltonian (10) describes a quantum transmission line with charge
discreteness, but it does not consider dissipation. On the other hand,
electrical resistance (Ohm law) is an intrinsic phenomenon in electrical
conduction. After Bateman's work [14], classical linear dissipation could be
studied in a Lagrangian way by consider a time-decaying exponential factor
multiplying the Langrangian function. From this, the classical Hamiltonian
could be written in the standard way. So, quantization becomes attainable
from the usual correspondence between position-momentum and its associated
operators [8,9,10]. In our case, a similar procedure could be implemented
and we obtain the time depending Hamiltonian:

\begin{equation}
\widehat{H}(t)=\sum_{l}\left\{ e^{-\frac{R}{L}t}\frac{2\hbar ^{2}}{Lq_{e}^{2}%
}\sin ^{2}\left( \frac{Lq_{e}}{2\hbar }\widehat{K}_{l}\right) +e^{\frac{R}{L}%
t}\frac{1}{2C}\left( \widehat{Q}_{l+1}-\widehat{Q}_{l}\right) ^{2}\right\} ,
\end{equation}
where the constant $R$ represents the resistance. With the above
Hamiltonian, and the Heinsenberg motion equations, we obtain for the charge
operator

\begin{equation}
\frac{d}{dt}\widehat{Q}_{l}=\frac{\hbar }{q_{e}L}e^{-\frac{R}{L}t}\sin
\left( \frac{Lq_{e}}{\hbar }\widehat{K}_{l}\right) ,
\end{equation}
and for the pseudo-current operator

\begin{equation}
\frac{d}{dt}\widehat{K}_{l}=\frac{1}{CL}e^{\frac{R}{L}t}\left( \widehat{Q}%
_{l+1}+\widehat{Q}_{l-1}-2\widehat{Q}_{l}\right) .
\end{equation}
From (19) and (20), the equation for the variation of the charge becomes

\begin{equation}
L\frac{d^{2}}{dt^{2}}\widehat{Q}_{l}=-R\frac{d}{dt}\widehat{Q}_{l}+\frac{1}{C%
}\left( \cos \left( \frac{Lq_{e}}{\hbar }\widehat{K}_{l}\right) \right)
\left( \widehat{Q}_{l+1}+\widehat{Q}_{l-1}-2\widehat{Q}_{l}\right)
\end{equation}
which in the formal limit $q_{e}\rightarrow 0$ becomes the usual one
incorporating a resistance $R$ in every cell of the transmission line. So,
the time depending Hamiltonian (18) describes the dynamics of a quantum
transmission line with charge discreteness and resistance.

\[
\]

\begin{section}*{ Conclusions and discussions}
\end{section}

We have proposed a quantum Hamiltonian for a transmission line composed of a
periodic array of inductances and capacitances with charge discreteness ((8)
or (10)). To construct this Hamiltonian we have used the charge discreteness
procedure proposed in reference [1] for a $LC$ circuit. In our case, the
corresponding Hilbert space is given by the tensorial product of this one of
every cell, corresponding to a $LC$ circuit. In the particular basis of
one-energy per site, or cell, we have found the spectrum of the line (17).

Note that charge discreteness produces nonlinear terms in the equation of
motion in the transmission line (11,12). The incorporation of electrical
resistance was considered by using the Caldirola-Kanai theory (section V).

As a final remark we note that disorder systems are usually studied in Solid
State Physics [15] and it is well-known that localization of states could
exist. In our case, we believe that disorder can be incorporated in the line
by consider for instance every inductance as a random quantity. In this
case, it seems interesting to study the role of disorder and the
nonlinearity due to charge discreteness. On the other hand, it is known that
decoherence effects break localization [16,17] then the role of environment
and resistance on the transmission line must also break localization.

\[
\]

{\bf Acknowledgments:} Discussion about the role of charge discreteness in
mesoscopic systems were carried-out with P. Orellana (Universidad
Cat\'{o}lica del Norte). Useful discussion were carried-out at the II
International Workshop on Disordered Systems (Arica, UTA 2000). This work
was supported by FONDECYT (Project 1990443).

\[
\]

\[
{} 
\]

\end{document}